\documentclass[twocolumn,showpacs,superscriptaddress,amsmath,aps]{revtex4}

\usepackage{amssymb,mathrsfs,bm,color,times}
\usepackage{graphicx,color}
\usepackage{CJK}
\usepackage{bm}
\usepackage[hypertex]{hyperref}
\usepackage{amsmath}
\usepackage{appendix}

\newcommand{\be}{\begin{equation}}
\newcommand{\ee}{\end{equation}}
\newcommand{\bea}{\begin{eqnarray}}
\newcommand{\eea}{\end{eqnarray}}
\newcommand{\bsube}{\begin{subequations}}
\newcommand{\esube}{\end{subequations}}

\newcommand{\Eq}[1]{Eq.\,(\ref{#1})}

\newcommand{\dg}{\dagger}
\newcommand{\la}{\langle}
\newcommand{\ra}{\rangle}

\newcommand{\beq}{\begin{equation}}
\newcommand{\eeq}{\end{equation}}
\newcommand{\beqn}{\begin{eqnarray}}
\newcommand{\eeqn}{\end{eqnarray}}

\newcommand{\nl}{\nonumber \\}

\newcommand{\bsub}{\begin{subequations}}
\newcommand{\esub}{\end{subequations}}

\begin{document}

\title{Gradual partial-collapse theory for ideal nondemolition
longitudinal-readout of qubits \\ in circuit QED}

\author{Wei Feng}
\email{fwphy@tju.edu.cn}
\affiliation{Center for Joint Quantum Studies and Department of Physics, Tianjin University,
Tianjin 300072, China}
\author{Cheng Zhang}
\affiliation{Department of Physics, Beijing Normal University,
Beijing 100875, China}
\author{Zhong Wang}
\affiliation{Department of Physics, Beijing Normal University,
Beijing 100875, China}

\author{Lupei Qin}
\affiliation{Center for Joint Quantum Studies and Department of Physics, Tianjin University,
Tianjin 300072, China}

\author{ Xin-Qi Li}
\email{xinqi.li@tju.edu.cn}
\affiliation{Center for Joint Quantum Studies and Department of Physics, Tianjin University,
Tianjin 300072, China}
\affiliation{Department of Physics, Beijing Normal University,
Beijing 100875, China}

\date{\today}

\begin{abstract}
The conventional method of qubit measurements in circuit QED
is employing the dispersive regime of qubit-cavity coupling,
which results in an approximated scheme of quantum nondemolition (QND) readout.
This scheme becomes problematic
in the case of strong coupling and/or strong
measurement drive, owing to the so-called Purcell effect.
A recent proposal by virtue of {\it longitudinal coupling}
suggests a new scheme to realize
fast, high-fidelity and {\it ideal QND} readout of qubit state.
The aim of the present work is twofold:
{\it (i)}
In parallel to what has been done in the past years for the dispersive readout,
we carry out the gradual partial-collapse theory for this recent scheme,
in terms of both the quantum trajectory equation
and quantum Bayesian approaches.
The partial-collapse weak measurement theory is useful for such as
the measurement-based feedback control and other quantum applications.
{\it (ii)}
In the physical aspect, we construct the joint qubit-plus-cavity
entangled state under continuous measurement
and present a comprehensive analysis for the quantum efficiency,
qubit-state purity, and signal-to-noise ratio in the output currents.
The combination of the joint state and the quantum Bayesian rule
provides a generalized scheme of cavity reset
associated with the longitudinal coupling,
which can restore the qubit to a quantum pure state
from entanglement with the cavity states, and thus
benefits the successive partial-collapse measurements after qubit rotations.
\end{abstract}


\maketitle
\section{Introduction}

The superconducting circuit quantum electrodynamics (cQED) architecture
has been a fascinating platform for quantum information processing
and for quantum measurement and control studies
\cite{Bla04,Sch04,Mooij04,Sch08,Pala10,DiCa13,Mar11,Dev13a,DiCa12,Dev13,Sid13,Sid15,Sid12,
Joh10,Joh12,DiV14B}.
For quantum measurement of qubits, which is a central ingredient of
many applications, the {\it dispersive coupling regime} is often exploited.
In this regime, the qubit and microwave cavity are off resonance,
and the Jaynes-Cummings coupling reduces to
an effective qubit-state dependent shift of the cavity frequency.
Then, a microwave drive to the cavity will result in
qubit-state dependent coherent state of the cavity field
and a homodyne detection of the field quadrature
can reveal information of the qubit state.

Actually, for qubit measurements in cQED,
this dispersive coupling readout is so far the typical and standard scheme
\cite{Bla04,Sch04,Mooij04,Sch08,Pala10,DiCa13,Mar11,Dev13a,DiCa12,Dev13,Sid13,Sid15,Sid12,
Joh10,Joh12,DiV14B}.
However, the dispersive scheme is approximate,
being applicable only for weak drive and weak coupling.
If either the measurement drive amplitude $\epsilon_m$
or the decay rate $\kappa$ of the cavity photons are increased,
the qubit will suffer state flip,
with a rate given by $\gamma_p=(\epsilon_m/\Delta)^2\kappa$,
where $\Delta$ is the detuning between the qubit energy and cavity frequency.
This is the so-called Purcell effect extensively discussed
in literature \cite{Bla11,Kor14,Wil16,Gam09,Sid12P,Reed10},
which results in the dispersive scheme
being not quantum nondemolition (QND).
One may notice other schemes of qubit measurements in cQED, e.g.,
the Josephson-bifurcation-amplifier readout \cite{Sid05,Sid06,Sid07,Lup07}
and the more recent studies of
quantum trajectories of superconducting qubit undergoing
homodyne-heterodyne detection of fluorescence
\cite{Huard14,Huard16,JorHua16,Jor17}.
However, the former is also nonideal QND while the latter is simply not.

In a recent work by Blais {\it et al.} \cite{Bla15-L},
an ideal-QND readout scheme for qubits in circuit QED
was proposed via parametric modulation of the
{\it longitudinal coupling} between the qubit and cavity field
\cite{Ker13,Nak15,Bla07,Bla09,Ric16,Koch07}.
This longitudinal-coupling scheme has been employed more recently
to study the quantum dynamics of simultaneously measured
noncommuting observables of a superconducting qubit \cite{Sid16}.
Differing from the (approximated) dispersive scheme,
the longitudinal coupling
acts as a qubit-state-dependent drive on the cavity.
That is, by performing the parametric modulation for the
longitudinal coupling (in microwave frequency regime),
the cavity field will evolve also into qubit-state-dependent coherent state
and a homodyne detection of the field quadrature
can reveal the qubit state information.
The analysis carried out in Ref.\ \cite{Bla15-L} showed that
the new protocol can outperform the dispersive scheme to achieve,
for instance, faster, high-fidelity, and ideally QND readout.
Moreover, further combination with using a squeezed microwave field
can even better improve the signal-to-noise ratio,
such as achieving the Heisenberg-limit scaling behavior
with the number of measuring photons \cite{Bla15-L,DiV14,Blai15B,Sid18}.

On the other aspect, concerning with quantum measurement and its applications,
a particular contemporary interest is the type of continuous weak measurement,
i.e., a continuous real-time monitoring of environment
with stochastic measurement records.
Despite the noisy backaction onto the measured state,
the stochastic evolution of the measured state, i.e., the so-called
quantum trajectory (QT), can be faithfully tracked \cite{WM09,Jac14}.
Therefore, this type of partial-collapse weak measurements can be useful
for quantum feedback \cite{Sid12,WM09,Jac14},
generating pre- and postselected quantum ensembles
to improve quantum state preparation, smoothing, and high-fidelity readout
and developing novel schemes of quantum metrology
\cite{Jord08,Li15,Li17,Mo13,Sid15,Mo14,Ts09,Wis15}.

For continuous weak measurements, the most celebrated theory
is the quantum trajectory equation (QTE),
as broadly applied in quantum optics
and quantum control studies \cite{WM09,Jac14}.
However, in some cases, a larger step state update scheme
known as quantum Bayesian approach can be more efficient,
by using the accumulated output currents
\cite{Sid12,Kor99,Kor11,Kor16,Li14,Li16}.
The Bayesian approach also allows for efficient and analytical studies
for interesting problems such as quantum weak values \cite{Jord08,Li15,Li17},
quantum state smoothing, quantum metrology \cite{Mo13,Sid15,Mo14,Ts09,Wis15}, etc.

So far, the partial-collapse weak measurement studies in cQED
(either QTE or Bayesian approach) were largely carried out
for dispersive readout
\cite{DiCa13,Mar11,Dev13a,DiCa12,Dev13,Sid13,Sid15,Sid12,Kor11,Kor16,Li14,Li16}.
The absence of similar investigations for the promising
longitudinal readout scheme motivates thus our present work.
As mentioned above for Ref.\ \cite{Bla15-L},
under {\it identical} (thus weak) coupling condition,
the longitudinal scheme is intrinsically faster than the dispersive readout
in concern with the projective measurement.
More importantly, the most essential point
is that the longitudinal coupling itself
is of exact quantum nondemolition while the effective dispersive coupling
is approximated under a weak coupling regime.
The natural immunity from the Purcell flip
allows possible strong longitudinal coupling
to achieve much faster readout speed.
For continuous partial-collapse measurement, this allows
faster information gain, which can benefit such as
the gained-information-based feedback to combat against decoherence.

In the present work, we carry out the gradual partial-collapse measurement theory
associated with the longitudinal readout scheme.
The work is organized as follows.
In Sec.\ II, we introduce the longitudinal coupling model
and present the ``original" QTE with the
cavity-photon degrees of freedom included,
conditioned on continuous homodyne detection of the cavity field quadrature.
In Sec.\ III, following Ref.\ \cite{Gam08}
while putting the details of derivation in the appendix,
we present the result of
{\it effective} QTE which contains only the qubit degrees of freedom.
Based on the result of the effective QTE and in particular
the constructed qubit-plus-cavity entangled state,
we present a comprehensive analysis for the quantum efficiency,
qubit-state purity, and signal-to-noise ratio in the output currents,
in comparison with the dispersive readout.
We further construct the quantum Bayesian rule in Sec.\ IV
and present an extended discussion on the issue of cavity reset in Sec.\ V.
We summarize the work in Sec.\ V with additional discussions.

\section{Model and Formulation}

In solid-state circuit QED, the standard scheme of qubit measurements
is the dispersive readout protocol \cite{Bla04,Sch04,Mooij04}.
Similar to its quantum optics counterpart
(i.e., the atomic cavity-QED system),
the most natural coupling between the superconducting qubit
and the resonator cavity is the Jaynes-Cummings Hamiltonian,
or, being a little bit more general
(without making rotating-wave approximation),
the so-called {\it transversal} coupling given by
$H_x=g_x \sigma_x (a+a^{\dg})$.
Here $a^{\dagger}$ ($a$) and $\sigma_x$ are respectively
the creation (annihilation) operator of the cavity photon
and the quasi spin operator for the qubit.
The so-called {\it dispersive regime} is defined by the criterion
that the detuning between the cavity frequency ($\omega_r$)
and qubit energy ($\omega_q$), $\Delta=\omega_r-\omega_q$,
is much larger than the coupling strength $g_x$.
Under this, the coupling Hamiltonian can be approximated by
$H_x\simeq \chi \sigma_z a^{\dagger}a$,
with $\chi=g_x^2/\Delta$.
Therefore, associated with this dispersive coupling Hamiltonian,
it is clear that the qubit states $|e\ra$ and $|g\ra$ will
change, respectively, the cavity frequency by $\pm\chi$.
Under measurement drive, the cavity field will evolve
into qubit-state-dependent coherent states and a dyne-type
quadrature detection can provide the qubit-state information.

In this work, following Ref.\ \cite{Bla15-L},
we consider an alternative coupling
between the qubit and cavity field, given by
\bea
H_z= g_z \sigma_z (a+ a^{\dagger})
\eea
In contrast to the {\it transversal coupling} which leads to an approximate
dispersive coupling Hamiltonian as shortly mentioned above,
this new type of coupling is termed as {\it longitudinal coupling},
with the subscript $z$ here to mark it.
This interaction Hamiltonian has been employed to discuss the realization
of multiqubit gates \cite{Ker13,Nak15,Bla07,Bla09,Ric16}.
In particular, in the superconducting circuits,
e.g., a flux or transmon qubit coupled to an inductor-capacitor ($LC$) oscillator \cite{Ker13,Nak15},
the longitudinal coupling can emerge from the mutual inductance
between the flux-tunable qubit and the oscillator.
Other examples include the general approach developed in Ref.\ \cite{Bla09}
based on a transmon qubit \cite{Koch07},
and the realization based on a transmission-line resonator \cite{Bla15-L}.

Let us now consider the parametric modulation of the longitudinal
coupling strength at the resonator frequency $\omega_r$,
i.e., $g_z(t)=\bar{g}_z+\tilde{g}_z\cos(\omega_r t)$.
In the rotating frame with respect to the free Hamiltonian
$H_0=\omega_r a^\dagger a +\frac{\omega_q}{2} \sigma_z$,
the interaction Hamiltonian reads \cite{Bla15-L}
\begin{equation}\label{H_eff}
H_z= \frac{1}{2} \tilde{g}_z \sigma_z ( a+ a^{\dagger} ) \,.
\end{equation}
In obtaining this result, all the terms of fast oscillations
have been neglected, including the $\bar{g}_z$ term.
It is clear that this modulation of the longitudinal coupling strength
plays a role of {\it qubit-state-dependent drive} to the cavity.
Instead of the dispersive readout
where the qubit-state-dependent shift of the cavity frequency is employed,
this alternative method can as well
result in qubit-state-dependent coherent state of the cavity field.
Via a homodyne quadrature detection of the cavity field,
the qubit state can be inferred.

As analyzed in detail in Ref.\ \cite{Bla15-L},
the longitudinal coupling scheme
has remarkable advantages to improve the qubit readout.
That is, it can lead to a faster, high-fidelity,
and ideally QND qubit readout with a simple reset mechanism.
Even with a conservative modulation, the qubit-state-dependent shift
of the cavity field can be easily distinguished.
Owing to the ideal QNDness of the coupling, i.e.,
being ideally immune to the measurement-caused qubit-state transition,
this scheme is free from the limitation of critical photon numbers.

The quadrature of the cavity field is detected as usual
by a homodyne measurement.
The measurement result can be expressed as \cite{WM09,Jac14}
\bea\label{Iphi}
I(t)=\sqrt{\kappa}\langle
ae^{-i\varphi}+a^\dag e^{i\varphi}  \rangle_{\varrho(t)}  +\xi(t) \,,
\eea
where $\varphi$ is the phase of the local oscillator
in the homodyne detection setup,
$\kappa$ is the leaky rate of the cavity photons,
and $\xi(t)$ satisfies the ensemble-average
properties of $E[\xi(t)]=0$ and $E[\xi(t)\xi(t')]=\delta(t-t')$.
The quantum average $\langle \cdots \rangle_{\varrho(t)}$
is defined by $\langle \cdots \rangle_{\varrho(t)}
={\rm Tr}[(\cdots)\varrho(t)]$, with $\varrho(t)$
the qubit-cavity conditional state
given by the quantum trajectory equation \cite{WM09,Jac14}
\begin{eqnarray}\label{QTE}
 \dot{\varrho} = -i[H_z,\varrho]
    +\kappa\mathcal{D}[a]\varrho
   +\sqrt{\kappa}\mathcal{H}[ae^{-i\varphi}] \varrho \xi(t) \;,
\end{eqnarray}
where the Lindblad superoperator is defined as
$\mathcal{D}[a]\varrho=a\varrho a^{\dagger}
-\frac{1}{2}\{a^{\dagger}a,\varrho \}$,
and the unraveling superoperator as
$\mathcal{H}[\bullet]\varrho = (\bullet)\varrho+\varrho (\bullet)^{\dg}
-\mathrm{Tr}\{[(\bullet)+(\bullet)^{\dg}]\varrho\}\varrho$.

In experiments, $\xi(t)$ in \Eq{QTE} is obtained from the output current
by using \Eq{Iphi}, while calculating the average $\la \cdots\ra_{\varrho(t)}$
in \Eq{Iphi} needs the knowledge of $\varrho(t)$ solved from \Eq{QTE}.
It is clear that this job is almost intractable
if the cavity photon number is large.
It would be thus desirable to establish an effective QTE
involving only the degrees of freedom of the qubit.
This can be done by applying the so-called polaron transformation
to eliminate the degrees of freedom of the cavity photons \cite{Gam08}.

\section{Quantum Trajectory Equation} \label{S-PT}

The basic idea of the polaron transformation is performing
a qubit-state-dependent displacement to the cavity field,
shifting it to a new vacuum state.
This allows us to simplify the procedures of tracing the cavity states
and obtain compact results for the equation of motion of the
(stochastic) reduced state of the qubit and the output current.

The polaron transformation is designed as \cite{Gam08}
\begin{eqnarray}\label{P}
P(t)=\Pi_e D[\alpha_e(t)]+\Pi_g D[\alpha_g(t)]  \,.
\end{eqnarray}
$D[\alpha_{e,g}(t)]$ are, respectively,
the displacement operators of the cavity fields
corresponding to the qubit states $|e\ra$ and $|g\ra$.
The specific form reads
$D[\alpha]=e^{\alpha a^\dagger-\alpha^*a}$.
The qubit-state dependence in the polaron transformation
is characterized by the projection operators of the qubit states
$\Pi_j=|j\rangle\langle j|$ $(j=e,g)$.
Under the parametric modulation of the longitudinal coupling
and in the presence of leakage of the cavity photons
(with rate $\kappa$),
the qubit-state-dependent evolution of the cavity field is governed by
\begin{eqnarray}\label{alphat}
\dot{\alpha}_{e,g}(t)=\mp i \tilde{g}_z/2-\kappa\alpha_{e,g}(t)/2 \,.
\end{eqnarray}
Starting to drive (modulate) the cavity from a vacuum,
the solution of the cavity field simply reads
\begin{eqnarray}\label{alpha-result}
\alpha_{e,g}(t)=\mp i\frac{\tilde{g}_z}{\kappa}(1-e^{-\kappa t/2}) \,.
\end{eqnarray}
It should be noted that the solutions of the cavity fields solved here
for the longitudinal readout are purely imaginary
and have different signs,
whereas for the dispersive readout the complex amplitudes
of the cavity states corresponding to $|e\ra$ and $|g\ra$
differ in the sign of real parts, but have the same imaginary parts \cite{Li14}.
As we will see in the following, this difference will result in
a different choice of the local oscillator's phase in the homodyne measurement
in order to optimize the information gain of
$|e\ra$ and $|g\ra$ of the qubit.
That is, for dispersive readout, $\varphi=0$;
while for the longitudinal readout, $\varphi=\pi/2$.
It is also this difference of $\alpha_{e}(t)$ and $\alpha_{g}(t)$
that will result in different rates
which fully characterize the effective QTE and the output current,
after eliminating the degrees of freedom of the cavity photons.

With the canonical transformation introduced above,
one can transform the qubit-cavity joint state
$\varrho^P(t)=P^{\dg}\varrho(t) P$,
and as well the two sides of \Eq{QTE}.
We may expand the transformed state in the qubit and
the photon-number (Fock) basis states,
$\varrho^P(t)=\sum^{\infty}_{n,m=0} \sum^{}_{i,j=e,g}
\varrho_{n,m;i,j}^P(t) |n,i\ra  \la m,j|$.
Owing to the cavity field shifted to a new vacuum, essentially,
only the zero-photon component will survive in this expansion.
However, nontrivial dynamics will result in excited states
of the cavity appearing in the transformed QTE.
Our interest is the {\it reduced} state of the qubit,
which is defined by tracing the cavity states
from the untransformed joint state,
$\rho(t)={\rm Tr}_R[P \varrho^P(t) P^{\dg}]$.
This establishes a connection between
$\rho_{ij}$ and $\varrho_{n,m;i,j}^P$.
Accordingly, we are able to obtain the equation of motion of $\rho(t)$,
from the ones of $\varrho_{n,m;i,j}^P(t)$
and by taking into account the fact that
only the $n,m=0$ components survive in the shifted state.
Following the procedures outlined above, we obtain
(see the appendix for more details)
\begin{eqnarray}\label{PAME}
  \dot{\rho} &=&   
   \Gamma_d(t) \mathcal{D}[\sigma_z]\rho
  -\sqrt{\Gamma_{ci}(t)} \mathcal{H}[\sigma_z]\rho \xi(t)  \nl
&& -i \sqrt{\Gamma_{ba}(t)}\, [\sigma_z,\rho]\xi(t)   \,.
\end{eqnarray}
Here the various rates read
\begin{subequations}\label{rates}
\begin{align}
& \Gamma_d(t)= \frac{\tilde{g}_z}{2} |\beta(t)|   \,, \\
& \Gamma_{ci}(t) =\frac{\kappa}{4} |\beta(t)|^2\cos^2(\varphi-\theta_\beta) \,, \\
& \Gamma_{ba}(t) = \frac{\kappa}{4}|\beta(t)|^2\sin^2(\varphi-\theta_\beta) \,.
\end{align}
\end{subequations}
In these results, we introduced
$\alpha_e(t)-\alpha_g(t)=\beta(t)=|\beta(t)|e^{i\theta_{\beta}}$.
Similarly, applying the same technique of transformation
to the calculation of the output current $I(t)$,
i.e., ${\rm Tr}[(\bullet)\varrho(t)]
= {\rm Tr}[P(\bullet)P^{\dg}\varrho^P(t)]$, we obtain
\begin{eqnarray} \label{It-2}
I(t)=-2\sqrt{\Gamma_{ci}(t)}\langle\sigma_z\rangle_t+\xi(t) \,.
\end{eqnarray}
In deriving this result, the fact that the shifted cavity field is a vacuum
should be used.

Formally, \Eq{PAME} is the same as the QTE for dispersive readout.
Similar results are also obtained in Ref.\ \cite{Sid16}.
Compared with the dispersive readout, a formal difference
only exists in the overall decoherence rate $\Gamma_d$ which reads, there,
$\Gamma_d=\chi\mathrm{Im}[\alpha_g(t)\alpha^*_e(t)]$.
However, the behaviors of the measurement rates
and the resultant consequences differ between both schemes,
owing to the different evolution of the cavity fields.
That is, in dispersive readout,
the qubit-state-dependent cavity fields are given \cite{Li14} by
$ \alpha_{e(g)}(t)= \bar{\alpha}_{e(g)}
(1-e^{\mp i\chi t-\kappa t/2})$, where
$ \bar{\alpha}_{e(g)} =-i\epsilon_m/(\pm i\chi + \kappa/2)$.
In dispersive readout,
the microwave driving amplitude $\epsilon_m$ corresponds to
the coupling modulation strength $\tilde{g}_z$ in the longitudinal scheme;
and moreover, there is an additional parameter $\chi$, i.e.,
the dispersive coupling strength between the qubit and the cavity field.
It is largely the difference of this solution
from the one given by \Eq{alpha-result}
for the longitudinal coupling
that results in the different behaviors to be shown in Fig.\ 1
associated with the following detailed analysis.

\subsection{Quantum efficiency and purity}

Briefly speaking,
$\Gamma_{ci}$ is the rate of {\it coherent information} gain,
i.e., inferring qubit state $|e\ra$ or $|g\ra$.
$\Gamma_{ba}$ characterizes the backaction of the measurement
not associated with information gain of the qubit state,
but on qubit-level fluctuations.
$\Gamma_{d}$ is the overall decoherence rate, after ensemble average
over large number of quantum trajectories.
The sum of the former two rates,
$\Gamma_m=\Gamma_{ci}+\Gamma_{ba}$, is the total measurement rate.
If $\Gamma_m=\Gamma_d$, the measurement is ideally quantum limited,
with quantum efficiency $\eta=\Gamma_m/\Gamma_d=1$.
Otherwise, if $\Gamma_m<\Gamma_d$, the measurement is not {\it ideal},
implying some information loss.

%

In Fig.\ 1(a), we plot the {\it transient} quantum efficiency,
defined as $\eta=\Gamma_m(t)/\Gamma_d(t)$,
of the longitudinal readout against its counterpart of the dispersive scheme,
by setting $\epsilon_m=\tilde{g}_z$ and taking several different $\chi$.
We find that, as the cavity field approaches to the steady state,
$\eta\rightarrow 1$ for both readout schemes,
which seemingly implies quantum-limited measurements.
However, this is only true in the sense
of {\it transient differential gain} of qubit-state information.
Starting the measurement over $(0,t)$
with a pure state of the qubit,
the resultant state of the qubit is given by \Eq{PAME},
which is no longer a pure state.
Actually, conditioned on the outcomes of the homodyne detection,
the qubit-plus-cavity joint state
can be expressed as \cite{Li14}
\bea\label{PsiT}
\Psi(t) =  c_1(t)|e\ra|\alpha_e(t)\ra
+ c_2(t) e^{i\Phi(t)} |g\ra|\alpha_g(t)\ra \,,
\eea
where the evolution of $c_1(t)$ and $c_2(t)$
corresponds to information gain of the qubit states,
and the (random) phase $\Phi(t)$ stems from
qubit energy fluctuations caused by measurement backaction.
The qubit state given by \Eq{PAME} corresponds to the one obtained
by tracing the cavity degrees of freedom from this entangled state.
This leads to a decoherence factor for the reduced state of qubit
given by \cite{Li14}
\bea\label{D-t}
D(t)=|\la\alpha_e(t)|\alpha_g(t)\ra|
=e^{-2\int^t_0 d\tau [\Gamma_d(\tau)-\Gamma_m(\tau)]} \,.
\eea
In general, this factor is smaller than unity,
implying a {\it degradation of purity} of the qubit state.

In Fig.\ 1(a), we also find that, at the early stage of partial-collapse measurement,
the quantum efficiency $\eta$ of the longitudinal readout
is better than the dispersive scheme.
However, this does not necessarily imply a higher {\it purity}
for the qubit state. In contrast, as shown in Fig.\ 1(b),
the corresponding purity of the qubit state, characterized by $D(t)$,
is lower than the dispersive readout.
The reason for this behavior is that
the {\it faster readout of the longitudinal scheme} is associated with
larger measurement rates $\Gamma_d(t)$ and $\Gamma_m(t)$.
Then, from \Eq{D-t}, we have a smaller purity factor $D(t)$,
owing to the larger difference of $\Gamma_d(t)-\Gamma_m(t)$.
However, this is not a serious problem for the longitudinal scheme.
When we introduce a fast reset procedure to the cavity,
the qubit can be disentangled with the cavity from \Eq{PsiT},
resulting thus in an ideal pure state for the qubit.
We will discuss this point in more detail near the end of next section.

With some surprise, we notice that in Fig.\ 1(a),
in the intermediate stage of dispersive readout,
it is possible to have $\eta>1$; i.e.,
the measurement rate $\Gamma_m$ is larger than
the associated decoherence rate $\Gamma_d$.
This is, in general, impossible.
From a general consideration of quantum system
coupled to the environment (with continuum of energy spectrum),
the continuous monitoring and measurement of the environment
has the consequence of unraveling the usual master equation.
However, in this case, the unraveling rate ($\Gamma_m$)
must be smaller than or at most equal to the decoherence rate ($\Gamma_d$).
Otherwise, consider starting with a pure state:
If $\Gamma_m>\Gamma_d$, the resultant state
(inferred from the measurement outcome) will be
"super-pure," being unphysical with the off-diagonal elements
of the density matrix violating the basic inequality of a physical state.
In the circuit QED system, however, this anomaly of the
transient differential rates
does not violate the requirement of a physical state.
Based on the structure of \Eq{PsiT} and the expression \Eq{D-t},
the transient $\Gamma_m(t)>\Gamma_d(t)$ simply means that
the purity (coherence) of the qubit state has recovered some amount.
This is interesting but possible.
Indeed, in Fig.\ 1(b), the plot of $D(t)$ supports this understanding.

\begin{figure}[!htbp]
  \centering
  \includegraphics[width=5.5cm]{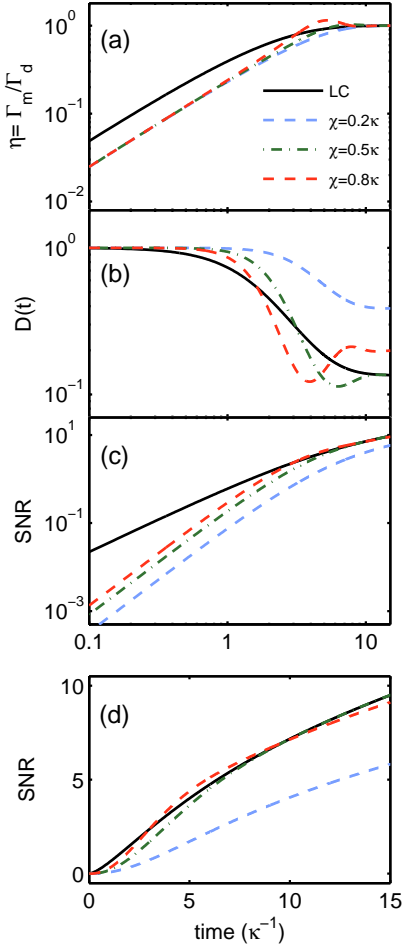}
  \caption{
Comparison between the longitudinal coupling (LC) readout
and the conventional dispersive readout scheme
(with several different coupling strengths,
defined from $\chi\sigma_z a^{\dg}a$).
In panels (a) and (b) we plot, respectively,
the transient quantum efficiency $\eta=\Gamma_m/\Gamma_d$,
and the purity degradation factor
$D(t)=|\la \alpha_e(t)|\alpha_g(t) \ra|$,
as a result of tracing out the degrees of freedom
of the cavity photons.
In panels (c) and (d), the same data of the signal-to-noise ratio (SNR)
are plotted, for different visualization purposes. }
\end{figure}

\subsection{Signal-to-noise ratio}

Based on the key results of Eqs.\ (\ref{PAME})-(\ref{It-2}),
it is also convenient to carry out another type of assessment,
i.e., the signal-to-noise ration (SNR) of the output currents.
From \Eq{It-2}, we can obtain the accumulated current over $(0,\tau)$ as
$Q=\int_0^{\tau} dt I(t)$.
In particular, corresponding to qubit states $|e\ra$ and $|g\ra$,
the accumulated currents are, respectively, $Q_e$ and $Q_g$.
Owing to the Gaussian noise $\xi(t)$,
$Q_e$ and $Q_g$ are random variables,
satisfying Gaussian distributions
with the distribution centers at
$\bar{Q}_{e(g)}=\mp \, 2\int_0^{\tau} \sqrt{\Gamma_{ci}(t)}dt$,
and with a variance (half-width) of $D_e=D_g=\sqrt{\tau}$.
Based on this picture, the SNR
of the measurement can be defined as
${\rm SNR}=|\bar{Q}_e-\bar{Q}_g|/(D_e+D_g)$.
Choosing $\varphi=\theta_\beta$, the results presented
in Ref.\ \cite{Bla15-L} are simply recovered as
\bea
{\rm SNR}_{(L)} =\tilde{g}_z
\sqrt{\frac{8\tau}{\kappa}}\left[1-\frac{2}{\kappa \tau}
(1-e^{-\frac{\kappa \tau}{2}})\right] \,,
\eea
for the longitudinal readout, and
\bea \label{SNR2}
{\rm SNR}_{(D)}
=\epsilon_m \sqrt{\frac{8\tau}{\kappa}}
\left[1-\frac{2}{\kappa \tau}(1-\cos(\frac{\kappa \tau}{2})
e^{-\frac{\kappa \tau}{2}})\right] \,,
\eea
for the dispersive readout.
Here, the optimal condition $\chi=\kappa/2$
has been assumed for the dispersive readout,
which ensures a maximal SNR in steady state of the cavity field
under given $\epsilon_m$ and $\kappa$.

As a preliminary comparison, i.e., neglecting the Purcell effect
in the dispersive readout, we set $\epsilon_m=\tilde{g}_z$.
From the above analytic results, we see that both SNR are identical
at steady state.
We see also from Figs.\ 1(c) and 1(d) that, indeed,
the condition of $\chi=\kappa/2$ can optimize
the SNR at steady state in the dispersive readout.
In the result of \Eq{SNR2}, the cosine term
may have effect in the transient process.
Indeed, we see from Fig.\ 1(d) that, for the choice of $\chi=0.8 \kappa$
as an example, the SNR of the dispersive readout can exceed
that of the longitudinal readout
in certain intermediate stages of the measurement process.

However, under equivalent parameter conditions,
in the short time regime,
the SNR of the longitudinal readout is better than
the dispersive readout,
just like the quantum efficiency shown in Fig.\ 1(a).
This merit may benefit the short-time
partial collapse measurement and the
measurement-based feedback control.

\section{Quantum Bayesian Rule} \label{S-Bay}

The effective QTE obtained above is much more
efficient than the original \Eq{QTE} which
contains the degrees of freedom of the cavity photons.
However, as illustrated in recent cQED experiments \cite{Dev13,Sid13,Sid15,Sid12},
the quantum Bayesian approach is an alternative convenient method
to update the quantum state based on accumulated output currents
over relatively larger time steps.
The quantum Bayesian rule allows also more efficient and analytical studies
for some interesting problems such as quantum weak values,
quantum smoothing, and quantum estimate \cite{Jord08,Li15,Li17,Mo13,Sid15,Mo14,Ts09,Wis15}.

In this section, we establish the quantum Bayesian rule
for the longitudinal readout
using the method developed in Ref.\ \cite{Li16}.
In the representation of the qubit basis states $|e\ra$ and $|g\ra$,
\Eq{PAME} is rewritten as
\begin{eqnarray}
\dot{\rho}_{ee}&=& -4\sqrt{\Gamma_{ci}}\rho_{ee}\rho_{gg}\xi   \,,   \label{pQTE-1a}\\
\dot{\rho}_{eg}&=& -2\left[ \Gamma_d\rho_{eg}
-\sqrt{\Gamma_{ci}}\langle\sigma_z\rangle\rho_{eg}\xi
+ i\sqrt{\Gamma_{ba}}\rho_{eg}\xi  \right] \,.  \nl
  \label{pQTE-1b}
\end{eqnarray}
Following Ref.\ \cite{Li16}, we establish the Bayesian rule
by straightforwardly integrating the stochastic differential equations.
However, special care is required.
While \Eq{PAME}, or the above equations, can work well
for numerical simulations based on the present It\'o calculus form,
an analytic solution should be obtained by
integrating its counterpart of the converted Stratonovich form.

We summarize the main results as follows.
First, for the diagonal elements, we have
\bea\label{BR-cQED-b}
\rho_{ii}(\tau) = \rho_{ii}(0) \, P_i (\tau)/ {\cal N}(\tau) \,,
\eea
where the index $i$ denotes $e$ and $g$,
and the normalization factor reads
${\cal N}(\tau)=\rho_{ee}(0)P_e(\tau) +\rho_{gg}(0) P_g(\tau)$.
Importantly, in contrast to the usual simple Gaussian function,
the {\it prior distributions} of the output currents
corresponding to qubit $|e\ra$ and $|g\ra$ are
instead {\it Gaussian functionals} given by
\bea\label{P12}
P_{i}(\tau)
= \exp\left\{-\la [I(t)-\bar{I}_{i}(t)]^2 \ra_{\tau} / (2V) \right\} \,.
\eea
Here we introduced the notations
$\la \bullet \ra_{\tau}=\tau^{-1}\int^{\tau}_{0} dt \, (\bullet)$
and $\bar{I}_{e(g)}(t)=\mp 2\sqrt{\Gamma_{ci}(t)}$.
The distribution variance $V$ simply reads $V=1/\tau$.

Second, the off-diagonal element is obtained as
\bea\label{BR-cQED-a}
&& \rho_{eg}(\tau) = \rho_{eg}(0)
   \left[\sqrt{P_e(\tau)P_g(\tau)}/{\cal N}(\tau) \right]  \nl
&& ~~~~~~ \times D(\tau) \,
\exp\left[-i\Phi(\tau)\right] \,.
\eea

The factor in the square brackets is associated with the
updated change of the diagonal elements of the qubit state.
Importantly, involved in this result for the off-diagonal element,
there are two more factors:
\bea
D(\tau)&=& \exp \left\{-2\int_0^{\tau}dt [\Gamma_d(t)
-\Gamma_m(t)]\right\} \,,   \label{BR-factors-a}\\
\Phi(\tau)&=& 2\int_0^{\tau} dt \sqrt{\Gamma_{ba}(t)}\,I(t) \,. \label{BR-factors-c}
\eea
Being closely related with the discussion in Sec.\ III A,
the first factor, $D(\tau)$,
solved here from the quantum trajectory \Eq{PAME},
accounts for the {\it purity degradation} of the qubit state
after tracing the cavity degrees of freedom
from the entangled qubit-plus-cavity state, \Eq{PsiT}.
Actually, it can be proved \cite{Li14} that
$D(\tau)=|\la \alpha_e(\tau)|\alpha_g(\tau)\ra|$,
which reveals, very directly,
the physical meaning of this factor.

The second one, $\Phi(\tau)$, is a random phase factor,
as also appeared in the entangled state of \Eq{PsiT}.
In contrast to the back-action owing to {\it information gain}
of the qubit state $|e\ra$ or $|g\ra$,
which can influence the diagonal elements or alters
$c_1(t)$ and $c_2(t)$ in \Eq{PsiT},
the phase factor $\Phi$ accounts for the backaction
of the measurement process (via the cavity photons)
on the energy-level fluctuation of the qubit.
Actually, the output current $I(t)$ carries also information
of the photon's duration in the cavity.
Via the ac-Stark effect, the cavity photons affect the qubit energy levels.
This consideration has been fully employed by Korotkov in Ref.\ \cite{Kor11}
and this type of stochastic phase fluctuations has been addressed particularly
in the framework of quantum trajectory equation for single-qubit state
and multiqubit parity measurements \cite{Gam08,Joh10,Joh12,DiV14B}.
In order to faithfully track the stochastic evolution
of the qubit state, one must take into account the effect of the random phase $\Phi(t)$.
In contrast to the usual reasoning of information loss,
this random fluctuation of the qubit energy levels is purity preserving
in the single realizations of continuous measurement, but does not cause decoherence.

In short, we may remind the readers that
when the gain of information on the qubit state is maximized
(by optimizing the local oscillator's phase),
the backaction associated with $\Gamma_{ba}$ [and thus the phase
factor $\Phi(t)$] vanishes completely.
Conversely, if a different phase of the local oscillator is used
and both $\Gamma_{ci}$ and $\Gamma_{ba}$ are nonzero,
the output current contains information on both kinds of backaction.

\section{Cavity Reset}

As briefly discussed in Sec.\ III A, the factor $D(t)$
implies that the partially collapsed qubit state
is not quantum mechanically pure.
Actually, from \Eq{PsiT}, we know that the qubit is entangled
with the cavity states $|\alpha_{e}(t)\ra$ and $|\alpha_{g}(t)\ra$.
This is a typical feature for some mesoscopic detectors,
where part degrees of freedom of the detector remain
in the state description.
A well-known example is the single-electron transistor
used as a charge-state detector \cite{set-1,set-2},
where the degrees of freedom
of the central dot or island cannot be eliminated in general.

However, in the present cQED setup, the cavity field is well structured.
After a partial-collapse measurement over $(0,t)$,
let us consider to perform a fast (shorter than $\kappa^{-1}$)
inverse displacement at the moment $t$ to the cavity field,
based on the same coupling Hamiltonian of \Eq{H_eff}
but inverting the phase of the modulation,
i.e., the action of
$ U_{\rm cav}(t+\delta t, t)|_{\delta t\to 0}
= e^{-i \tilde{G}_z\delta t[\sigma_z(a+a^{\dg})]}$
on the cavity state.
Here we denote the coupling amplitude by $\tilde{G}_z$
(rather than $\tilde{g}_z$), implying a strongly pulsed drive.
We may re-denote the result of \Eq{alpha-result}
as $\alpha_{e,g}(t)=\mp i a$.
Now, if we make $\tilde{G}_z\delta t=-a$
(noting the particular sign opposite to
the measurement drive $\tilde{g}_z$),
we actually {\it disentangle} the qubit from
the qubit-plus-cavity entangled state, as follows:
\bea\label{joint-state-1}
&\Psi(t)&  =  c_1(t)|e\ra|\alpha_e(t)\ra
+ c_2(t) e^{i\Phi(t)} |g\ra|\alpha_g(t)\ra  \nl
&\rightarrow&
\Psi(t+0^+) =  [ c_1(t)|e\ra + c_2(t) e^{i\Phi(t)} |g\ra] |0\ra    \,,
\eea
where $|0\ra$ denotes the cavity vacuum.
Note that this is essentially the same procedure of {\it cavity reset}
discussed in Ref.\ \cite{Bla15-L},
but a generalization from the projectively collapsed state
$|\bar{\alpha}_e\ra$ or $|\bar{\alpha}_g\ra$
to the partially collapsed state of a superposition
of $|\alpha_e(t)\ra$ and $|\alpha_g(t)\ra$.
Therefore, after this qubit-state-dependent displacement,
{\it we completely restore the desired qubit state},
which is quantum mechanically pure,
with now {\it no degradation of purity}
owing to disentangling with the cavity states.

Unfortunately, if we introduce further qubit transition (e.g., unitary operation),
the situation becomes rather complicated.
In the context of dispersive readout, this issue has been discussed
in connection with the quantum trajectory equation
by Gambetta {\it et al.} \cite{Gam08}
and with the quantum Bayessian approach by Korotkov \cite{Kor16}.
Only in the so-called bad-cavity limit,
will the cavity states associated with $|e\ra$ and $|g\ra$
 rapidly relax to $|\bar{\alpha}_e\ra$ and $|\bar{\alpha}_g\ra$.
This makes the transient dynamics of the cavity field have negligible effect.
As a result, each new step evolution affected by the measurement
is updated by using the Bayesian inference, based on measurement rates
determined by the stationary states of the cavity field.
Beyond the bad-cavity limit,
this will be an open problem for further investigations.
For the longitudinal readout scheme, the difficulty may be partly solved
by a fast reset procedure (before the qubit rotation)
as described by \Eq{joint-state-1}.

\section{Summary and Discussions}

In parallel to what have been done for the dispersive readout, we carried out
a study on the gradual partial-collapse weak measurement theory
for a recently proposed genuinely QND readout scheme
by virtual of longitudinal coupling in circuit QED.
Keeping in mind that this scheme has the advantage of being free from
the Purcell-effect-induced flip of qubit state,
our results are expected to have sound application potential for
such as feedback control and other quantum technologies
associated with partial-collapse weak measurements.

The theory was constructed in terms of both the quantum-trajectory equation
and quantum Bayesian approaches.
In particular, we construct the joint qubit-plus-cavity entangled state
under continuous measurement.
Combining this {\it joint state} with the {\it quantum Bayesian rule}
can result in a generalized scheme for cavity reset,
which can improve the multiple partial-collapse measurements
in the presence of qubit rotations.
Based on the time-dependent measurement rates,
we carried out comprehensive studies on the key figures of merit
including transient quantum efficiency, qubit state purity,
and signal-to-noise ratio of the measurement.

The figures of merit of these quantities
indicate that the longitudinal readout is somehow superior to the dispersive scheme,
e.g., as shown in Fig.\ 1, having better quantum efficiency and signal-to-noise ratio
for short time partial-collapse measurement.
However, the really significant point is that, the longitudinal scheme
is a genuine QND measurement, allowing
strong drive (with large $\tilde{g}_z$)
but suffering no flip of the qubit state when the cavity field
exceeds some critical photon numbers $n_{\rm crit}$.
So one can safely increase the modulation amplitude $\tilde{g}_z$
to improve the SNR.
In sharp contrast, in the dispersive readout,
increasing either the drive amplitude $\epsilon_m$
or the decay rate of the cavity photons
will induce the non-QNDness caused by Purcell effect,
with a flip rate of the qubit state given by
$\gamma_p=(\epsilon_m/\Delta)^2\kappa$.
In practice, this Purcell-effect-induced flip of qubit state should be
avoided in quantum measurement.
The longitudinal readout scheme has great advantage in this essential aspect.
However, the dispersive coupling is reduced from
the most natural Jaynes-Cummings coupling Hamiltonian,
being thus more straightforwardly implemented.
This is the reason that so far the dispersive coupling readout
scheme is typically used in practice.

The other advantage of the longitudinal coupling is that it allows us to
design a more efficient method to reset the cavity field to vacuum.
This can extend the applicability of the effective QTE
and Bayesian approach beyond the bad-cavity limit. Actually,
the reset scheme is rooted in the insight gained from the constructed
{\it qubit-cavity joint state} during the partial-collapse measurement.
This entangled joint state may shine also light on developing
quantum trajectory theory under generic parameter conditions
in the presence of qubit rotations,
which remains so far an open question
within the framework of effective QTE
after eliminating the cavity degrees of freedom.

Finally, we mention that, in Ref.\ \cite{Bla15-L}
and in a few recent publications \cite{DiV14,Blai15B,Sid18},
squeezing of microwave-frequency fields
has been discussed for qubit measurement of circuit quantum electrodynamics.
The ability to couple squeezed fields to superconducting qubits can enable
faster measurement and allow encoding of more information per photon,
thus resulting in Heisenberg-limit-scaling behavior.
In a very recent experiment \cite{Sid18},
based on stroboscopic longitudinal coupling and squeezing field injected,
enhanced signal-to-noise ratio for qubit measurement has been observed.
Extension of qubit readout using squeezed field from the projective measurement
to partial-collapse weak measurement is of great interest
and will be a subject for further investigations.  \\
\\
\\
{\it Acknowledgements}---
This work was supported by the National Key Research
and Development Program of China (No. 2017YFA0303304)
and the National Natural Science Foundation of China (No. 11675016).\\
\\
\appendix 
\section*{appendix: Derivation of \Eq{PAME}}

Closely following the notation used in Ref.\ \cite{Gam08},
let us introduce the two quadratures of the cavity field
\begin{subequations}
\begin{eqnarray}\label{quad}
 && I_\varphi=\frac{1}{2} \left(a e^{-i\varphi}+a^\dagger e^{i\varphi}\right) \,,  \\
 && Q_\varphi=-\frac{i}{2} \left(a e^{-i\varphi}-a^\dagger e^{i\varphi}\right) \,.
\end{eqnarray}
\end{subequations}
Then, we rewrite \Eq{QTE} as
\begin{eqnarray}\label{STE}
\dot{\varrho}&=&-i[H_z,\varrho]
  +\kappa\mathcal{D}[a]\varrho+\gamma_1\mathcal{D}[\sigma_-]\varrho
  +\frac{\gamma_2}{2}\mathcal{D}[\sigma_z]\varrho \nl
  &+&\sqrt{\kappa}\mathcal{M}[2I_\varphi]\varrho\xi
   +i\sqrt{\kappa}[Q_\varphi,\varrho]\xi \,.
\end{eqnarray}
The newly introduced superoperator reads
$\mathcal{M}[x]\varrho=\{x,\varrho\}/2- \la x \ra \varrho$,
where the quantum average follows the usual meaning
$\la x \ra=\mathrm{Tr}(x\varrho)$.
To be a little more general (not considered in this work),
we also included that the external environment caused
(not caused by the measurement itself) relaxation and dephasing
with respective rates $\gamma_1$ and $\gamma_2$.

Below, starting with \Eq{STE},
we apply the so-called polaron transformation to eliminate
the cavity degrees of freedom to obtain \Eq{PAME}, i.e.,
the effective QTE with only the qubit degrees of freedom.
For the sake of clarity, we treat first the deterministic terms,
then the stochastic ones.

\subsection{The deterministic part}

The polaron transformation is constructed as \Eq{P},
with the respective displacement of the cavity fields
given by \Eq{alpha-result}.
Performing the transformation $\varrho^P=P^{\dagger}\varrho P$
(and similarly for other operators),
after simple algebra, we obtain
 \begin{eqnarray}\label{PME}
 &&\dot{\varrho }^P =P^{\dagger} \dot{ \varrho} P
 +\dot{P}^{\dagger} \varrho P+ P^{\dagger} \varrho \dot{P} \nonumber\\
 &&=-i[H_{z}^{P},\varrho^{P}]
  +\kappa\mathcal{D}[a^{P}]\varrho^{P}+\gamma_1\mathcal{D}[\sigma_-^{P}]\varrho^{P}
  +\frac{\gamma_2}{2}\mathcal{D}[\sigma_z^{P}]\varrho^{P} \nonumber\\
 && + \, [\dot{P}^{\dagger} P,\varrho^{P}]  \,.
\end{eqnarray}
In this equation the displaced operators read
\begin{subequations}
\begin{eqnarray}
 &&a^{P}=a+\Pi_\alpha=a+(\beta/2) \sigma_z \,,  \\
 &&\sigma_z^{P}=\sigma_z  \,,    \\
 &&\sigma_-^{P}=\sigma_{-}D[\beta] \,.
\end{eqnarray}
\end{subequations}
We may note that $D[\beta]$ is the displacement operator
defined in the main text.
In rewriting the first equation, we noticed that
$\Pi_\alpha=\alpha_e\Pi_e+\alpha_g\Pi_g$
and $\beta=\alpha_e-\alpha_g$.
Substituting them into \Eq{PME}, we obtain
 \begin{eqnarray}\label{TME}
 &&\dot{\varrho }^P = \kappa\mathcal{D}[a]\varrho ^P
 +  \left( \frac{\kappa\beta}{2}
 [\sigma_z,\varrho ^P]a^\dagger + {\rm H.c.}\right)  \nonumber\\
  &&+ \gamma_1\mathcal{D}[\sigma_{-} D(\beta)]\varrho ^P
+\frac{1}{2}(\gamma_2+\kappa|\beta|^2/2)\mathcal{D}[\sigma_z]\varrho ^P  \,. \nl
\end{eqnarray}
Note that the reduced state of the qubit should be obtained from
$\rho(t)={\rm Tr}_R[\varrho(t)]={\rm Tr}_R[P \varrho^P(t) P^{\dg}]$.
Expanding the {\it transformed state} in the qubit and
the photon-number (Fock state) basis,
$\varrho^P(t)=\sum^{\infty}_{n,m=0} \sum^{}_{i,j=e,g}
\varrho_{n,m;i,j}^P(t) |n,i\ra  \la m,j|$,
we obtain
\begin{eqnarray} \label{RDM}
&& \rho= \mathrm{Tr}_R[P\varrho^P P^\dagger]      \nonumber\\
&&    =\sum_n[\varrho _{n,n;e,e}^P|e\rangle\langle e|
+\varrho _{n,n;g,g}^P|g\rangle\langle g|]  \nonumber\\
&&    +\sum_{n,m}[\lambda_{n,m;m,n}|e\rangle\langle g|
+\lambda_{m,n;n,m}^*|g\rangle\langle   e|] \,,
\end{eqnarray}
where $\lambda_{n,m;p,q}\equiv \varrho ^P_{n,m;e,g} d_{p,q}
\exp(-i\rm Im[\alpha_g\alpha_e^*])$,
with $d_{p,q}=\langle p|D(\beta)|q\rangle$
being the matrix element of the displacement operator in the photon-number basis.
Based on \Eq{TME}, the following set of equations for the matrix elements
in the above expansion can be derived:
\begin{subequations}
\begin{eqnarray}\label{mq-ee}
&& \dot{\varrho}_{n,m;e,e}^P = -[\gamma_1+\frac{\kappa}{2}(n+m)]\varrho _{n,m;e,e}^P \nl
&& +\kappa \sqrt{(n+1)(m+1)}\varrho _{n+1,m+1;e,e}^P     \,,
\end{eqnarray}
\begin{eqnarray}\label{mq-gg}
&& \dot{\varrho}_{n,m;g,g}^P = -\frac{\kappa}{2}(n+m)\varrho _{n,m;g,g}^P \nl
&& + \gamma_1 \sum_{p,q} \varrho_{p,q;e,e}^P d_{n,p}d_{m,q}^*    \nl
&& + \kappa \sqrt{(n+1)(m+1)}  \varrho_{n+1,m+1;g,g}^P   \,,
\end{eqnarray}
\begin{eqnarray}\label{mq-pq}
&& \dot{\lambda}_{n,m;p,q}= -[\frac{\gamma_1}{2}+\gamma_2+2\Gamma_d
+\frac{\kappa}{2}(n+m)]\lambda_{n,m;p,q} \nl
&&+\kappa\beta\sqrt{m+1}\lambda_{n,m+1;p,q}-\kappa\beta^*\sqrt{n+1}\lambda_{n+1,m;p,q} \nl
&&+\kappa\sqrt{(n+1)(m+1)}\lambda_{n+1,m+1;p,q}+\dot{\beta} \sqrt{p}\lambda_{n,m;p-1,q} \nl
&&-\dot{\beta} \sqrt{q}\lambda_{n,m;p,q-1}   \,.
\end{eqnarray}
\end{subequations}
In the last eqation, the decoherence rate reads
$\Gamma_d=\frac{\tilde{g}_z}{2} |\beta|$.

Then, we obtain the equations for the reduced density matrix elements
of the qubit state as follows:
\begin{subequations}\label{DMelements}
\begin{eqnarray}
&& \dot{\rho}_{ee} =\sum_n  \dot{\varrho}^P_{n,n;e,e}=-\gamma_1 \rho_{ee}^P \,,  \\
&& \dot{\rho}_{gg} =\sum_n  \dot{\varrho}^P_{n,n;g,g}=\gamma_1 \rho_{ee}^P \,, \\
&& \dot{\rho}_{eg} =\dot{\lambda}_{0,0;0,0}=\left(-\frac{\gamma_1}{2}
-\gamma_2-2\Gamma_d \right) \rho_{eg} \,.
\end{eqnarray}
\end{subequations}
In achieving these results, we employed the fact that
the displaced state $\varrho^P$ has a cavity vacuum
so only the zero-photon component survives.
However, coupling dynamics can result in the first excited state
of the cavity, as indicated in the right-hand side of Eq.\ (A7).
In a basis-free form, we re-express the results as
\begin{equation}\label{ME}
  \dot{\rho} = \gamma_1\mathcal{D}[\sigma_-]\rho
  + (\gamma_2/2+\Gamma_d) \mathcal{D}[\sigma_z]\rho \,.
\end{equation}

\subsection{Including the stochastic part}

To make the derivation simpler, we consider the following {\it linear} QTE
for the unnormalized state $\bar{\varrho}$:
\begin{eqnarray}
\dot{\bar{\varrho}}&=& \mathcal{L} \bar{\varrho}
  + \sqrt{\kappa }\bar{\mathcal{M}}[2I_\varphi]\bar{\varrho}\xi
   +i\sqrt{\kappa}[Q_\varphi,\bar{\varrho}]\xi  \,.
\end{eqnarray}
Here we denote the deterministic part as $\mathcal{L} \bar{\varrho}$,
while the linear unraveling superoperator simply reads
$\bar{\mathcal{M}}[x]\bar{\varrho}=\frac{1}{2} \{x,\bar{\varrho}\}$.
Following precisely the same treatment for the deterministic part, we obtain
\begin{eqnarray}
\dot{\bar{\varrho}}^P&=& \mathcal{L} \bar{\varrho}^P
  + \sqrt{\kappa}|\beta| \cos(\theta_\beta-\varphi) \bar{\mathcal{M}}[\sigma_z]
  \bar{\varrho}^P\xi     \nl
 && + \sqrt{\kappa }(a e^{-i\varphi} \bar{\varrho}^P
 +\bar{\varrho}^Pa^\dagger e^{i\varphi}) \xi   \nl
 && +\frac{i}{2}\sqrt{\kappa}|\beta| \sin(\theta_\beta-\varphi)
 [\sigma_z,\bar{\varrho}^P]\xi  \,.
\end{eqnarray}
For the matrix elements introduced in \Eq{RDM}, we have
\begin{subequations}
\begin{eqnarray}
&& \dot{\bar{\varrho}}_{n,m;e,e}^P = (\ref{mq-ee}) \nl
&&~~~ +\sqrt{\kappa}|\beta| \cos(\theta_\beta-\varphi) \bar{\varrho}^P_{n,m;e,e}\xi \nl
&&~~~ +\sqrt{\kappa} \sqrt{n+1}e^{-i\varphi}\bar{\varrho}^P_{n+1,m;e,e} \xi   \nl
&&~~~ +\sqrt{\kappa} \sqrt{m+1}e^{i\varphi}\bar{\varrho}^P_{n,m+1;e,e} \xi  \,,   \nl
\end{eqnarray}

\begin{eqnarray}
&& \dot{\bar{\varrho}}_{n,m;g,g}^P = {\rm r.h.s.~of~ Eq}.\ (\ref{mq-gg}) \nl
&&~~~ -\sqrt{\kappa}|\beta|
\cos(\theta_\beta-\varphi) \bar{\varrho}^P_{n,m;g,g}\xi \nl
&&~~~ +\sqrt{\kappa} \sqrt{n+1}e^{-i\varphi}\bar{\varrho}^P_{n+1,m;g,g} \xi   \nl
&&~~~ +\sqrt{\kappa} \sqrt{m+1}e^{i\varphi}\bar{\varrho}^P_{n,m+1;g,g} \xi   \,,  \nl
\end{eqnarray}
\begin{eqnarray}
&& \dot{\bar{\lambda}}_{n,m;p,q} = {\rm r.h.s.~of~ Eq}.\ (\ref{mq-pq}) \nl
&&~~~   +\sqrt{\kappa}\sqrt{n+1}e^{-i\varphi} d_{p,q} \bar{\varrho}^P_{n+1,m;e,e} \xi\nl
&&~~~ +\sqrt{\kappa}\sqrt{m+1}e^{i\varphi}  d_{p,q} \bar{\varrho}^P_{n,m+1;e,e} \xi\nl
&&~~~ +i\sqrt{\kappa}|\beta| \sin(\theta_\beta-\phi) \bar{\lambda}_{n,m;p,q} \xi  \,.
\end{eqnarray}
\end{subequations}
Under the same spirit of obtaining Eqs.\ (A8a)-(A8c),
we obtain the following equations
for the reduced density matrix elements of the qubit state:
\begin{subequations}
\begin{eqnarray}
\dot{\bar{\rho}}_{ee}&=&\dot{\bar{\varrho}}_{0,0;e,e}^P
= {\rm r.h.s.~of~ Eq}.\ (A8a) \nl
&& + \sqrt{\kappa}|\beta|
\cos(\theta_\beta-\varphi) \bar{\rho}_{ee} \xi  \,, \nl
\end{eqnarray}
\begin{eqnarray}
\dot{\bar{\rho}}_{gg}&=&\dot{\bar{\varrho}}_{0,0;g,g}^P
= {\rm r.h.s.~of~ Eq}.\ (A8b)  \nl
&& -\sqrt{\kappa}|\beta|
\cos(\theta_\beta-\varphi) \bar{\rho}_{gg} \xi  \,, \nl
\end{eqnarray}
\begin{eqnarray}
\dot{\bar{\rho}}_{eg}&=&\dot{\bar{\lambda}}_{0,0;0,0}
={\rm r.h.s.~of~ Eq}.\ (A8c) \nl
&& + i\sqrt{\kappa}|\beta|
\sin(\theta_\beta-\varphi) \bar{\rho}_{eg} \xi   \,. \nl
\end{eqnarray}
\end{subequations}
Converting these results into a basis-free form
yields the following linear QTE
\begin{eqnarray}
  \dot{\bar{\rho}} &=& \mathcal{L} \bar{\rho}
    -2\sqrt{\Gamma_{ci}(t)} \bar{\mathcal{M}}[\sigma_z] \bar{\rho} \xi
 -i \sqrt{\Gamma_{ba}(t)}[\sigma_z,\bar{\rho}]\xi   \,.  \nl
\end{eqnarray}
The expressions of the measurement rates $\Gamma_{ci}(t)$
and $\Gamma_{ba}(t)$ are referred to \Eq{rates}.

Finally, we normalize the state via $\rho=\bar{\rho}/{\rm tr}(\bar{\rho})$
and obtain its equation of motion, i.e., the effective QTE
after eliminating the cavity photon degrees of freedom, as
\begin{eqnarray}
  \dot{\rho} &=& \mathcal{L} \rho
    -2\sqrt{\Gamma_{ci}(t)} \mathcal{M}[\sigma_z] \rho \xi(t)
 -i \sqrt{\Gamma_{ba}(t)}[\sigma_z,\rho]\xi(t)  \,.  \nl
\end{eqnarray}
This is the \Eq{PAME} we employed in the main text.


\begin{references}
\bibitem{Bla04}
A. Blais, R. S. Huang, A. Wallraff, S. M. Girvin,  and R. J. Schoelkopf,
Phys. Rev. A {\bf 69}, 062320 (2004).

\bibitem{Sch04}
A. Wallraff,  D. I. Schuster, A. Blais, L. Frunzio, R. S. Huang,
J. Majer, S. Kumar, S. M. Girvin,  and  R. J. Schoelkopf,
Nature (London) {\bf 431}, 162 (2004).

\bibitem{Mooij04}
I. Chiorescu, P. Bertet, K. Semba, Y. Nakamura,
C. J. P. M. Harmans,  and J. E. Mooij,
Nature (London) {\bf 431}, 159 (2004).

\bibitem{Sch08}
R. J. Schoelkopf, and S. M. Girvin,
 Nature (London) {\bf 451}, 664 (2008).

\bibitem{Pala10}
A. Palacios-Laloy, F. Mallet, F. Nguyen, P. Bertet, D. Vion, D. Esteve, and A. N. Korotkov,
Nat. Phys. {\bf 6}, 442 (2010).

\bibitem{DiCa13}
J. P. Groen, D. Rist\'e, L. Tornberg, J. Cramer, P. C. de Groot,
 T. Picot, G. Johansson, and L. DiCarlo,
Phys. Rev. Lett. {\bf 111}, 090506 (2013).

\bibitem{Mar11}
M. Mariantoni, H. Wang, R. C. Bialczak, M. Lenander, E. Lucero,
M. Neeley, A. D. O’Connell, D. Sank, M. Weides, J. Wenner, T. Yamamoto, Y. Yin,
J. Zhao, J. M. Martinis, and A. N. Cleland,
Nat. Phys. {\bf 7}, 287 (2011).


\bibitem{Dev13a}
P. Campagne-Ibarcq, E. Flurin, N. Roch, D. Darson,
P. Morfin, M. Mirrahimi, M. H. Devoret,
F. Mallet,  and B. Huard,
Phys. Rev. X {\bf 3}, 021008 (2013).

\bibitem{DiCa12}
D. Rist\'e, J. G. van Leeuwen, H. S. Ku,
K. W. Lehnert, and L.  DiCarlo,
Phys. Rev. Lett. {\bf 109}, 050507 (2012).

\bibitem{Dev13}
M. Hatridge, S. Shankar, M. Mirrahimi, F. Schackert, K. Geerlings,
T. Brecht, K. M. Sliwa, B. Abdo, L. Frunzio, S. M. Girvin,
R. J. Schoelkopf, and M. H. Devoret,
Science {\bf 339}, 178 (2013).
\bibitem{Sid13}  
K. W. Murch, S. J. Weber, C. Macklin, and I. Siddiqi,
Nature {\bf 502} (London), 211 (2013).
\bibitem{Sid15}
D. Tan, S. J. Weber, I. Siddiqi, K. Molmer, and K. W. Murch,
 Phys. Rev. Lett. {\bf 114}, 090403 (2015).
\bibitem{Sid12}
R. Vijay, C. Macklin, D. H. Slichter, S. J. Weber, K. W. Murch,
R. Naik, A. N. Korotkov, and I. Siddiqi,
Nature  (London) {\bf 490}, 77 (2012).
\bibitem{Joh10}  
L. Tornberg and G. Johansson, Phys. Rev. A {\bf 82}, 012329 (2010).
\bibitem{Joh12}
A. F. Kockum, L. Tornberg, and G. Johansson,
Phys. Rev. A {\bf 85}, 052318 (2012).
\bibitem{DiV14B}
L. Tornberg, Sh. Barzanjeh, and D. P. DiVincenzo,
Phys. Rev. A {\bf 89}, 032314 (2014).
%
%
\bibitem{Bla11}
F. Beaudoin, J. M. Gambetta, and A. Blais,
Phys. Rev. A  {\bf 84}, 043832 (2011).
\bibitem{Kor14}
E. A. Sete, J. M. Gambetta, and A. N. Korotkov,
Phys. Rev. B {\bf 89}, 104516 (2014).
\bibitem{Wil16}
L. C. G. Govia and F. K. Wilhelm,
Phys. Rev. A {\bf 93}, 012316 (2016).
\bibitem{Gam09}
M. Boissonneault, J. M. Gambetta, and A. Blais,
Phys. Rev. A {\bf 79}, 013819 (2009).
\bibitem{Sid12P}
D. H. Slichter, R. Vijay, S. J. Weber, S. Boutin,
M. Boissonneault, J. M. Gambetta, A. Blais, and I. Siddiqi,
Phys. Rev. Lett. {\bf 109}, 153601 (2012).
 \bibitem{Reed10}
M. D. Reed, B. R. Johnson, A. A. Houck, L. Dicarlo,
J. M. Chow, D. I. Schuster, L. Frunzio, and R. J. Schoelkopf,
Appl. Phys. Lett.  {\bf96}, 203110 (2010).




%
\bibitem{Sid05}  
I. Siddiqi, R. Vijay, F. Pierre, C. M. Wilson, M. Metcalfe,
C. Rigetti, L. Frunzio, and M. H. Devoret,
Phys. Rev. Lett. {\bf 93}, 207002 (2004);
I. Siddiqi, R. Vijay, F. Pierre, C. M. Wilson,
L. Frunzio, M. Metcalfe, C. Rigetti, R. J. Schoelkopf, M. H.
Devoret, D. Vion, and D. Esteve, {\it ibid.} {\bf 94}, 027005 (2005).
\bibitem{Sid06}
I. Siddiqi, R. Vijay, M. Metcalfe, E. Boaknin, L. Frunzio, R. J. Schoelkopf,
and M. H. Devoret, Phys. Rev. B {\bf 73}, 054510 (2006).
\bibitem{Sid07}
V. E. Manucharyan, E. Boaknin, M. Metcalfe, R. Vijay, I. Siddiqi,
and M. Devoret, Phys. Rev. B {\bf 76}, 014524 (2007).
\bibitem{Lup07}
A. Lupascu, S. Saito, T. Picot, P. C. de Groot,
C. J. P. M. Harmans, and J. E. Mooij, Nat. Phys. {\bf 3}, 119 (2007).

%
 \bibitem{Huard14}  
P. Campagne-Ibarcq, L. Bretheau, E. Flurin, A. Auff\`{e}ves, F. Mallet, and B. Huard,
Phys. Rev. Lett. {\bf112}, 180402 (2014).
 \bibitem{Huard16}
P. Campagne-Ibarcq, P. Six, L. Bretheau, A. Sarlette, M. Mirrahimi, P. Rouchon, B. Huard,
Phys. Rev. X {\bf6}, 011002 (2016).

\bibitem{JorHua16}
A. N. Jordan, A. Chantasri, P. Rouchon, and B. Huard,
Quantum Stud. Math. Found. 3, 237 (2016).

 \bibitem{Jor17}
M. Naghiloo, D. Tan, P. M. Harrington, P. Lewalle, A. N. Jordan, and K. W. Murch,
Phys. Rev. A {\bf 96}, 053807 (2017).




\bibitem{Bla15-L}   
N. Didier, J. Bourassa, and A. Blais,
Phys. Rev. Lett. {\bf 115}, 203601 (2015).
\bibitem{Ker13}
A. J. Kerman, New J. Phys. {\bf15}, 123011 (2013).
\bibitem{Nak15}
P. M. Billangeon, J. S. Tsai, and Y. Nakamura, Phys. Rev. B {\bf 91}, 094517 (2015).
\bibitem{Bla07}
A. Blais, J. Gambetta, A. Wallraff, D. I. Schuster,
S.M. Girvin, M. H. Devoret, and R. J. Schoelkopf,
Phys. Rev. A {\bf 75}, 032329 (2007).
\bibitem{Bla09}
J. Bourassa, J. M. Gambetta, A. A. Abdumalikov, O. Astafiev, Y. Nakamura, and A. Blais,
 Phys. Rev. A {\bf 80}, 032109 (2009).
\bibitem{Ric16}
S. Richer and D. Divincenzo, Phys. Rev. B {\bf 93}, 134501 (2016).
\bibitem{Koch07}
J. Koch, T. M. Yu, J. Gambetta, A. A. Houck, D. I. Schuster,
J. Majer, A. Blais, M. H. Devoret, S. M. Girvin,
and R. J. Schoelkopf, Phys. Rev. A {\bf 76}, 042319 (2007).

\bibitem{Sid16}
S. Hacohen-Gourgy, L.Martin, E. Flurin, V. V. Ramasesh,
K. B. Whaley, and I. Siddiqi,
Nature (London) {\bf 538}, 491 (2016)

\bibitem{DiV14}
Sh. Barzanjeh, D. P. DiVincenzo, and B. M. Terhal,
Phys. Rev. B {\bf 90}, 134515 (2014).
\bibitem{Blai15B}
N. Didier, A. Kamal, W. D. Oliver, A. Blais, and A. A. Clerk,
Phys. Rev. Lett. {\bf 115}, 093604(2015).
\bibitem{Sid18}
A. Eddins, S. Schreppler, D. M. Toyli, L. S. Martin, S. Hacohen-Gourgy,
L. C. G. Govia, H. Ribeiro, A. A. Clerk, and I. Siddiqi,
Phys. Rev. Lett. {\bf 120}, 040505(2018).


%
\bibitem{WM09}  
Wiseman, H. M. \&  Milburn, G. J. {\it Quantum Measurement
and Control} (Cambridge University Press, Cambridge,UK, 2009).
\bibitem{Jac14}
Jacobs, K. {\it Quantum Measurement Theory and
Its Applications} (Cambridge University Press, Cambridge,UK, 2014).

%
%

\bibitem{Jord08}
N. S. Williams and A. N. Jordan,
Phys. Rev. Lett. {\bf 100}, 026804 (2008).
\bibitem{Li15}
L. Qin,  P. Liang, and X. Q. Li,
Phys. Rev. A {\bf 92}, 012119 (2015).
\bibitem{Li17}
L. Qin, L. Xu, W. Feng, and X. Q. Li,
New J. Phys. {\bf 19}, 033036 (2017).

\bibitem{Mo13}
S. Gammelmark, B. Julsgaard, and K. Molmer,
Phys. Rev. Lett. {\bf 111}, 160401 (2013).

\bibitem{Mo14}
T. Rybarczyk, S. Gerlich, B. Peaudecerf, M.Penasa,B. Julsgaard,
K. Moelmer, S. Gleyzes, M. Brune, J.-M Raimond, S. Haroche,
and I. Dotsenko, arXiv:1409.0958.

\bibitem{Ts09}
M. Tsang, Phys. Rev. A {\bf 80}, 033840 (2009).
\bibitem{Wis15}
I. Guevara, and H. Wiseman,
Phys. Rev. Lett. {\bf 115}, 180407 (2015).

%
\bibitem{Kor99}
A. N. Korotkov, {\it Phys. Rev. B} {\bf 60}, 5737 (1999).
\bibitem{Kor11}
A. N. Korotkov, arXiv:1111.4016; see also A. N. Korotkov,
in {\it Quantum Machines: Measurement and Control of Engineered
Quantum Systems}, edited by M. Devoret, B. Huard, R.
Schoelkopf, and L. F. Cugliandolo (Oxford University Press,
New York, 2014), Vol. 96, Chap. 17.

\bibitem{Kor16}
A. N. Korotkov, Phys. Rev. A {\bf 94}, 042326 (2016).
\bibitem{Li14}
P. Y. Wang, L. P. Qin, and X.-Q. Li,
New J. Phys. {\bf 16}, 123047 (2014);
{\it ibid.} {\bf 17}, 059501 (2015).
\bibitem{Li16}
W. Feng, P. F. Liang, L. P. Qin, and X.-Q. Li,
Sci. Rep. {\bf 6}, 20492 (2016).

%


\bibitem{Gam08}
J. Gambetta, A. Blais, M. Boissonneault, A. A. Houck,
D. I. Schuster, and S. M. Girvin,
Phys. Rev. A {\bf77}, 012112 (2008).

\bibitem{set-1}
M. H. Devoret and R. J. Schoelkopf,
Nature (London) {\bf 406}, 1039 (2000).
\bibitem{set-2}
Yu. Makhlin, G. Sch\"on, and A. Shnirman,
Rev. Mod. Phys. {\bf 73}, 357 (2001).


\end{references}
\end{document}